\documentclass{PoS}

\PoS{PoS(LAT2005)196}

\usepackage{epsfig}
\usepackage{amsmath}
\title{Screening of sources in higher representations of SU(N) gauge 
theories at zero and finite temperature}

\ShortTitle{Sources in higher SU(N) representations }

\author{\speaker{Ferdinando Gliozzi}\\
    Dipartimento di Fisica Teorica,    Universit\`a di Torino and\\
INFN, sezione di Torino, Italy\\

        E-mail: \email{gliozzi@to.infn.it}}

\abstract{
Completion of the Svetitsky-Yaffe conjecture in some (2+1) dimensional
SU(N) gauge theories allows mapping Polyakov loops in higher
representations of the gauge group into suitable conformal operators of the
corresponding 2D CFT. As a consequence, the critical exponents of the 
correlators of these  Polyakov loops are determined. The
functional form of these correlators suggests a general Ansatz to
describe  the large-distance screening of  higher-representation sources 
at zero temperature in any space-time dimension. A generalised Wilson loop in 
which along part of its trajectory a source is converted in a gauge 
invariant way into higher representations with  same $N-$ality could  
be used to estimate the decay scale of the unstable strings.  }

\FullConference{XXIIIrd International Symposium on Lattice Field Theory\\

                 25-30 July 2005\\

                 Trinity College, Dublin, Ireland}

\newcommand{\Z}{\mathbb{Z}}

\newcommand{\uu}{\mathbb{I}}
\newcommand{\R}{{\cal R}}

\newcommand{\bra}{\langle}
\newcommand{\ket}{\rangle}

\newcommand{\tr}{{\rm tr}\,}
\newcommand{\Tr}{{\rm tr}^{~}}
\newcommand{\eq}{\begin{equation}}
\newcommand{\en}{\end{equation}}
\newcommand{\bea}{\begin{eqnarray}}
\newcommand{\ea}{\end{eqnarray}}

\begin{document}

\section{Introduction}
The confining phase of  $SU(N) $ gauge theory is characterised by a linear 
rising potential between static sources in the fundamental representation 
due to formation of a confining string joining these sources. It is natural 
to ask what is the potential between static  sources in higher 
representations. One faces two somewhat related problems. 

At $T=0$ two 
sources in a irreducible representation $\R$ and its conjugate $\bar{\R}$ 
give rise to the formation of a confining string with a string tension 
$\sigma_\R$. Most strings of this kind  are expected to be unstable:  if the 
representation $\R$ is built up of $j$ 
copies of the fundamental representation $f$, then $\sigma_\R$ should depend
only on the $N$-ality of $\R$, defined as $k_\R\equiv j$ (mod $N$),
 because all representations with same $k$ can be converted into each other 
by the emission of a proper number of soft gluons, therefore  
the heavier $\R-$strings are expected to decay into the string with smallest 
string tension within the same $N$-ality. However 
this expectation seems not supported by numerical experiments.
For a recent discussion on this subject see \cite{as}.

The other problem concerns  the critical behaviour of the Polyakov lines 
in arbitrary representations at $T_c$.
Universality arguments would, for continuous deconfining  transitions, place 
the finite-temperature $SU(N)$ gauge theory in the universality class 
of $\Z_N$ invariant spin models in one dimension less \cite{sy}. 
The spin operator is mapped into the Polyakov line in the fundamental
representation. What about the Polyakov lines in higher representations of 
$SU(N)$? There appears to be no room for independent exponents for these 
higher representations from the point of view of the abelian spin system.
Also such an expectation  seems not supported by numerical experiments
(See \cite{dh} and references therein). 

In this talk  I describe  a map between the operator product expansion (OPE) 
of the Polyakov operators in the gauge theory and the corresponding spin 
operators in the two-dimensional conformal field theory which describes the 
associated spin system at criticality which solves in some cases such an 
apparent puzzle and gives a hint to find a general solution of the related 
problem  at $T=0$. A  more detailed account of this approach  as well as a 
complete list of references  can be found in \cite{Gliozzi:2005dv}. 

\section{Polyakov loops at criticality}
\label{polyakov}
Consider a $2+1$ dimensional pure gauge theory undergoing a continuous 
deconfinement transition at the critical temperature $T_c$. The 
effective model describing the behaviour of Polyakov lines at 
finite $T$ will be  a
 two-dimensional spin model with a global symmetry group coinciding with 
the center of the gauge group. According to the Svetitsky and Yaffe 
 (SY) conjecture \cite{sy} such a 
spin model belongs in the same universality class of the original 
gauge theory. 

Using the methods of conformal field theory (CFT), the 
critical behaviour can be often determined 
exactly. For example, the critical properties of $2+1$ dimensional $SU(3)$ 
gauge theory at deconfinement coincide with those of the 3-state Potts model,

What is needed to fully exploit the predictive power of the SY conjecture 
is a mapping relating the physical observables of the gauge theory to the 
operators of the dimensionally reduced model..

The correspondence between the  Polyakov line 
in the fundamental representation
$f$ and the order parameter $\sigma$ of the spin model is the first entry  
in this mapping:
\eq
\Tr_f(U_{\vec{x}}) ~\sim~ \sigma(\vec{x})~, 
\en
$U_{\vec{x}}$ is the gauge group element  associated to the closed path 
wound   once around the periodic imaginary time at the point $\vec{x}$. 
It is now natural to ask what operators in the CFT correspond to
Polyakov lines in  higher representations. On the gauge side,
these can be obtained by a proper combination of products of Polyakov 
lines in the fundamental representation, using repeatedly   
the OPE of  Polyakov operators in the 
fundamental representation:
\eq
  \Tr_f(U_{\vec{x}})\; \tr_f(U_{\vec{y}})=
\sum_{\R\in f\otimes f}C_\R(\vert\vec{x}-\vec{y}\vert)
\,\Tr_{\R}(U_{(\vec{x}+\vec{y})/2})
+\dots
\en
where the coefficients $C_\R(r)$ are suitable functions (they become  powers
of $r$ at the critical point) and the dots represent the contribution of 
higher dimensional local operators. The important property of  this OPE
is that the local operators are classified according to the irreducible 
representations of $G$ obtained by the decomposition of the direct product 
of the representations of the two local operators in the left-hand side.

\par On the CFT side, we have a similar structure.  The order parameter
$\sigma$ belongs to an irreducible representation $[\sigma]$ of the Virasoro 
algebra  
and the local operators contributing to an OPE are 
classified according to the decomposition of the direct product of the 
Virasoro representations of the left-hand-side operators. This decomposition 
is known as the \emph{fusion algebra}.

In the case of three-state Potts model there is  a finite number of 
representations, listed here along with their scaling dimensions
\bea
{\uu}~({\rm identity});&~ \sigma,\,\sigma^+~({\rm spin~fields});~
\epsilon~({\rm energy});~\psi,\psi^+\\
x_{\uu}=0;& x_\sigma=\frac2{15};~~~~ x_\epsilon=
\frac4{15};~~~~~ x_\psi=\frac43~.
\label{potts}
\ea

The fusion rule we need is
\eq
[\sigma]\star[\sigma]=[\sigma^+]+[\psi^+]~;
\label{fusion3}
\en 

Comparison of the first equation with the analogous one of the gauge side
\eq
\{3\}\otimes\{3\}=\{\bar{3}\}+\{6\}~,
\en
 yields a new entry of the gauge/CFT correspondence
 \eq
\Tr_{\{6\}}(U_{\vec{x}}) ~\sim~ \psi^+(\vec{x})+
c\,\sigma^+(\vec{x})~,
\label{c6}
\en
where there are no a priori reasons for the vanishing of the 
coefficient $c$. Hence the Polyakov-Polyakov critical correlator
of the symmetric representation $\{6\}$ is expected to have the 
following general form in the thermodynamic limit
\eq
\bra\Tr_{\{6\}}(U_{\vec{x}})\;\Tr_{\{\bar{6}\}}(U_{\vec{y}})
\ket_{SU(3)}^{~}=
\frac{c_s}{r^{2x_\sigma}}+\frac{c_u}{r^{2x_\psi}}~,
\label{c66}
\en
 with $r=\vert\vec{x}-\vec{y}\vert$ and $c_s,c_u$ suitable coefficients.
Since $x_\sigma<x_\psi$, the second term drops off more rapidly than the 
first, thus at large distance this correlator behaves like that of 
the anti-symmetric representation $\{\bar{3}\}$ as expected also at zero 
temperature. 

A similar reasoning  can be applied to other  representations of 
$SU(3)$ and $SU(4)$, as it has been explicitly  worked out in
Ref. \cite{Gliozzi:2005dv}.

\section{Decay of unstable strings at zero temperature} 
\label{decay}

 The difficulty in observing  string 
breaking  or string decay with the Wilson loop seems to indicate nothing 
more than that it has a very small overlap with the broken-string or 
stable string state.
Why? being a general phenomenon which occurs for any gauge group, 
including $\Z_2$, in pure gauge models as well as in models coupled 
to whatever kind of matter, it requires a general explanation which should 
not depend on detailed dynamical properties of the model. 
A simple, general, explanation in the case of gauge models coupled 
with matter was proposed in  \cite{Gliozzi:2004cs} and can be easily 
generalised to the present case \cite{Gliozzi:2005dv} .

The general form of the Polyakov correlator in higher representations of 
$SU(3)$ found in Eq.s(\ref{c66})  suggests a simple  
Ansatz describing the asymptotic functional form
of the vacuum expectation value of a large, rectangular, Wilson loop in 
a higher representation $\R$ coupled to an unstable string which should 
decay into a stable  $k-$string
\footnote{For sake of simplicity we neglect the $1/r$ term in the potential 
which accounts for the quantum fluctuations of the flux tube.}
\eq 
\bra W_\R(r,t)\ket\simeq c_u\exp[-2\mu_{\R}^{~}(r+t)-\sigma_{\R}^{~} rt]+c_s
\exp[-2\mu_\R' (r+t)-\sigma_k^{~} rt]
\label{Ansatz} 
\en
The first term describes the typical area-law decay produced at intermediate 
distances by the unstable string with tension $\sigma_{\R}^{~}$. 
The second term is  instead the contribution expected 
by the stable $k-$string in which the $\R-$string decays. In the case 
of adjoint representation (zero $N-$ality) 
one has $\sigma_0=0$ and the perimeter term $\mu'_{adj}$ denotes
the mass of the lowest glue-lump. Eq.(\ref{Ansatz}) has to be understood as an 
asymptotic expansion  which approximates  $\bra W_\R(r,t)\ket$ when
$r,t>r_o$, where $r_o$ may be interpreted as the  scale where the confining 
string forms.

When $t$ and $r$ are sufficiently large, no 
matter how small $c_s$ is, the above Ansatz 
implies that at  long distances the stable string  
eventually prevails, since 
$\Delta\sigma\equiv\sigma_\R-\sigma_k>0$, hence the first
 term drops off more rapidly than the second and we have
\eq
V_\R(r)=V_s(r)=2\mu'_\R+\sigma_k^{~}r~,~~~ r>r_\R^{~}~,
\label{pots}
\en
where $V_\R(r)=-\lim_{t \to\infty}\frac1t\log\bra W_\R(r,t)\ket$ 
 is the static potential and $r_\R^{~}$ is the decay scale, given by  
\eq
r_\R^{~}=\frac{2\Delta \mu}{\Delta \sigma}\equiv2\frac{\mu'_\R-\mu^{~}_\R}
{\sigma^{~}_\R-\sigma^{~}_k}~.
\label{scale}
\en

In the case of 
zero $N-$ality the above equation yields the usual estimate of the adjoint 
string breaking scale.  
 Notice that the mass $\mu_\R^{~}$ and 
$\mu'_\R$ are not UV finite because of the additive  self-energy divergences, 
which cannot be 
absorbed in a parameter of the theory. However, these divergences should 
cancel in their difference, hence $r^{~}_\R$ is a purely dynamical scale, 
defined  for any non fully antisymmetric representation of $SU(N)$, which 
cannot be tuned by any bare parameter of the theory. When $r$ is less than t
he decay scale $r^{~}_\R$ Eq.(\ref{pots}) is no longer 
valid and has to be replaced by
\eq
V_\R(r)=V_u(r)=2\mu^{~}_\R+\sigma_\R^{~}r~,~~~ r_o<r<r_\R^{~}~.
\label{potu}
\en
Thus, the Ansatz (\ref{Ansatz}) describes the unstable string decay as a level
crossing phenomenon, as already observed in the string breaking. 
From a computational point of view it is very challenging  to 
check this Ansatz. As a matter of fact, only in 2+1 $SU(2)$ adjoint Wilson loop
\cite{krde} and in in 2+1 $\Z_2$-Higgs model \cite{Gliozzi:2004cs} the 
string  breaking has been convincingly demonstrated in this way.

\begin{figure}
\centering
\includegraphics[width=0.6\textwidth]{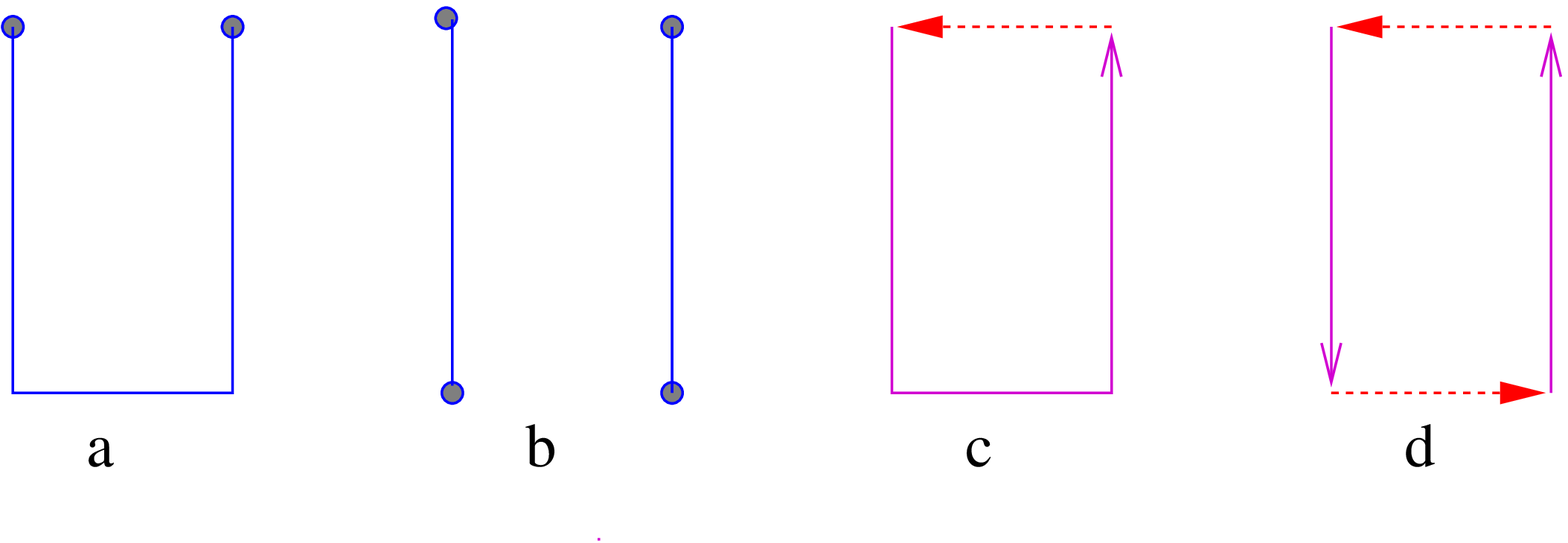} 
\caption{The operators depicted in 
$a$ and $b$ describe adjoint Wilson lines decaying in the vacuum. They 
are used to extract the adjoint static potential within the multichannel 
method. The mixed Wilson loops  $c$ and $d$ generalise these operators to 
the case of representations of non zero $N-$ality.} 
\label{Figure:1}
\end{figure}

\section{Mixed Wilson loops}
The winning move to easily observe the breaking of the adjoint string has been 
the proposal \cite{cm} of enlarging the basis of of the operators used to 
extract the potential.
Adjoint sources, contrarily to what happens in the case of fundamental 
representation, can form gauge invariant \emph{open} Wilson lines, like those 
depicted in Fig. \ref{Figure:1} $a$ and $b$, having a good overlap with 
the broken adjoint string state. This suggests  generalising such a 
construction to an excited irreducible representation $\R$ with non-vanishing 
$N$-ality $k$ by trying to construct operators as those depicted in 
Fig. \ref{Figure:1} $c$ and $d$  where  along one or more segments of the 
closed path $\gamma$ (the dashed lines in Fig.\ref{Figure:1}) the static 
source carries the quantum numbers of $\R$, 
while in the remaining path 
(solid lines) the source lies in the stable, fully anti-symmetric 
representation $k$.

To construct such a \emph{mixed Wilson loop} let us start by considering an 
arbitrary closed path $\gamma=u\,v$ made by the composition of two open paths
$u$ and $v$. Let $U$ and $V$ be the group elements associated with 
these two paths respectively. The associated standard Wilson loop is
\eq
W_f(\gamma)=\tr(UV)~.
\label{wo}
\en  
where the trace is taken, here and in the following, in the fundamental 
representation $f$  of the \emph{\sl non-abelian} gauge group $G$. We want to 
transform $W$ in a mixed Wilson loop in which the source along the path $u$ 
carries the quantum numbers of an
higher representation $\R$ belonging to the same $N-$ ality  of $f$.
To this end we perform the replacement 
\eq
U\to UQU^\dagger P U
\en  
where $P$ and $Q$ are the group elements associated to two small loops inserted
at the end points of the path $u$, in analogy with the ``clovers'' used 
in the construction of the glue-lump operator \cite{cm}.
  Along the path $u$ now propagates a  source belonging to the 
reducible representation $f\otimes f\otimes\bar{f} $. We have then to 
project on some irreducible component.
 To make a specific, illustrative example, let us consider the case of 
$SU(3)$, where we have $\{3\}\otimes \{3\}\otimes
\{\bar{3}\}=2\{3\}+\{\bar{6}\}+\{15\}$.
 We want to project on the $\{\bar{6}\}$ representation which has the same
triality of $\{3\}$. It is easy to find 
\eq
\begin{split}
W^{~}_{\{3\}\to\{\bar{6}\}}(\gamma)=\frac32\left[\tr(PUQU^\dagger)\,\tr(VU)-
\tr(PUVUQU^\dagger)\right]+\\
\frac34\left[\tr P\, \tr(VUQ)-\tr(VPUQ)+\tr (VPU)\,\tr Q-\tr P\tr(VU)\,\tr Q
\,\right]~.
\label{w36}
\end{split}
\en
Other examples and more details on the projections on 
the irreducible representations can be found in Ref.\cite{Gliozzi:2005dv}.
 \par
It is clear that the above construction can be generalised to any non-abelian 
group and, in particular, to any fully anti-symmetric representation 
of $N-$ality $k$ of $SU(N)$, which can be converted through the emission and 
the reabsorption of a glue-lump to an excited representation $\R$.
It is also clear that one can build up mixed Wilson loops of the 
type drawn in Fig.\ref{Figure:1} $c$ and $d$ that we denote respectively as 
$W^{~}_{k\to\R}(r,t)$ and $W^{~}_{\R\to k\to\R}(r,t)$
\par The static potential $V_\R(r)$ between the $\R$ sources and the 
decay of the 
associated unstable string into the stable $k-$string can then be extracted 
in the standard way from measurements of the matrix correlator
\eq
C(r,t)=\left(
\begin{matrix}
\bra W^{~}_{\R}(r,t)\ket &\bra W^{~}_{k\to\R}(r,t)\ket\\
\bra W^{~}_{k\to\R}(r,t)\ket&\bra W^{~}_{\R\to k\to\R}(r,t)\ket
\end{matrix}
\right)~,
\label{cor}
\en   
where $W^{~}_{\R}$ is the ordinary Wilson loop  when the whole 
rectangle $(r,t)$ is in the $\R$ representation. This is the 
generalisation of the multichannel method which has been  used successfully to 
observe the breaking of  the string in gauge theories coupled to matter 
fields as well as the adjoint string breaking.  In this way we  are confident 
that it will also possible to evaluate the decay scale $r_\R$.
 
One could also try to get a rough estimate of $r_\R$  through 
Eq.(\ref{scale}). Indeed the instability of the $\R-$string leads to 
conjecture that the vacuum expectation value of 
$W^{~}_{k\to\R}(\gamma)$ should behave asymptotically as 
\eq
\bra W^{~}_{k\to\R}(\gamma)\ket\propto e^{-\mu_\R^{~}\vert v\vert-
\mu_\R'\vert u\vert-\sigma^{~}_k A}~,
\en
where $A$ is the area of the minimal surface encircled by $\gamma$ and 
$\vert v\vert$ and $\vert u\vert$  are the lengths of the paths which carry 
 the quantum numbers of the representations $k$ and $\R$, respectively.
This seems the most effective way to estimate the quantity $\mu'_\R$
and therefore $r^{~}_\R$.

\end{document}